# Does IT Matter (Now)? A Global Panel Data Analysis of 7 Regions from 2018-2020 on Digitalization and Its Impact on Economic Growth

**Full research paper**

**Mahikala Niranga**
Faculty of Business, Law and Arts
Southern Cross University
Gold Coast, Australia
Email: mahi44n@gmail.com

**Darshana Sedera**
Faculty of Business, Law and Arts
Southern Cross University
Gold Coast, Australia
Email: darshana.sedera@scu.edu.au

**Golam Sorwar**
Faculty of Science and Engineering
Southern Cross University
Gold Coast, Australia
Email: golam.sorwar@scu.edu.au

## Abstract

There has been a long-running debate in Information Technology (IT) and economics literature about the contrary arguments of IT concerning digitalization and the economic growth of nations. While many empirical studies have shown a significant value of IT, others revealed a detrimental impact. Given the ambiguous results and anecdotal commentary on the increase in digitalization attributed to the COVID-19 global pandemic, this paper aims to explore the economic growth-digitalization nexus of 59 countries in 7 regions by employing correlation and regression analyses over the period 2018-2020. The findings indicate a positive relationship between economic growth and digitalization for both 'HIGH' and 'LOW' digitalized country categorization and regional assessment. Consistent with regional results, except for Northern Africa and Western Asia, and Sub-Saharan Africa regions, the remaining regions show a positive correlation and regression results. The findings of this study can be helpful in future prospective national IT and economic development policies.

# 1 Introduction

In 2003, Harvard scholar Nicholas G. Carr, brought a controversial argument regarding the value of IT in his article "IT Doesn't Matter," questioning the value of corporate IT spending and its relationship with superior financial outcomes (Carr 2003). Understandably, scholars responded to Carr's original assertion regarding the strategic worth of IT for a company and emphasised the details of his case for the broader macroeconomic benefits of the industry (Brown and Hagel 2003; Vennamaneni 2016). Over the last decades, the value of IT and economic performance have been well-established (Sedera and Lokuge 2017; Sedera and Lokuge 2019) and there is ample evidence as to how IT provides a competitive edge through the digitalization of work practices (Wamboye et al. 2015).

Due to the growing significance of digitalization, academics are examining the economic return of digitalization from many aspects (Fang et al. 2020). Numerous theoretical and empirical studies have been carried out to discover the solutions. The body of literature demonstrates that several theories acknowledge that digitalization is increasingly essential in accelerating economic growth and productivity (Bahrini and Qaffas 2019). Considering the purported 'hype of digitalization' spreading across the globe at present, especially attributing to the COVID-19 global pandemic, required us to re-visit this question again at a macro-level. For example, there has been a long-running debate in IT and economics literature about the positive and negative impacts on economic well-being, growth rates and productivity (Aleksandrova and Khabib 2021; Bahrini and Qaffas 2019; Carr 2003). While many empirical studies have shown that digital technology, particularly in developed countries, plays a positive and significant role in enhancing economic growth (Aleksandrova and Khabib 2021), other research revealed that digitalization has a detrimental impact on economic growth across various countries and regions of the world (Bahrini and Qaffas 2019; Krutova et al. 2021).

The COVID-19 pandemic has provided individuals, firms and nations with the chance to observe the effects of digitalization in a way that has never before been possible due to a pandemic of this magnitude (Fang et al. 2020). In 2019, 204 billion apps were downloaded; as of January 2020, 3.8 billion people actively used social media ( Alarifi et al. 2015; Kemp 2020). Especially with the lockdowns and remote-working required by the COVID-19 restrictions, there is a growing assumption that the whole world is moving fast in digitalization than in previous years (Frost and Duan 2020; Pan et al. 2020). It is likely to become even more critical for every region in the future as it seems to offer a vast potential to enhance productivity and shape economic growth (Alshubiri et al. 2022; Krutova et al. 2021). As of 2021, 67% of the global population had subscribed to mobile devices, of which 75% were smartphones, and the mobile industry contribution was 5% of GDP (GSMA 2022). As per the ambiguous results, this association between digitalization and economic growth is still open to investigation. There are still severe precise variances in results and gaps in the knowledge as no one has considered, especially with the changes that occurred during the COVID-19 pandemic. Given this, this paper aims to empirically investigate the economic growth-digitalization nexus of 59 countries in 7 regions by employing correlation and simple regression analyses over the period 2018-2020. To achieve this objective, the fundamental research question we explore is "*Does digitalization contribute to the economic growth of a country?*".

# 2 Literature Review

Digitalization has become the most visible change in recent decades, and it is expected to grow even more in the future as technology advances. Accordingly, digital work shows a new pathway for strategy makers of organizations in a holistic way (Amrollahi et al. 2020; Morton et al. 2020; Sedera 2006). Governments, individual households, and businesses can all adopt digital technologies to varying degrees (Aldhafiri et al. 2019; Sanchez-Riofrio et al. 2022). All of these activities together make up an economy's digitization index (ITU 2017), and it has been primarily evaluated through the digitalization of consumption (market digitalization) and digitalization of production, as described in earlier literature (Katz et al. 2020; Sanchez-Riofrio et al. 2022). The size of the digitalized market is referred to as market digitalization (e.g., number of people with broadband access, use of Internet and Information Communication Technology (ICT) skills) (Sanchez-Riofrio et al. 2022), while digitalization of production refers to firms' investments in digital technologies and services to improve transaction efficiency and market expansion (Katz et al. 2020).

Moreover, digitalization has gained central attention in previous research due to its substantial impact on economic growth (Bahrini and Qaffas 2019; Dedrick et al. 2003; Krutova et al. 2021). Academics and professionals that have concentrated on researching the association between digitalization and economic growth have become increasingly interested in the rapid global progress of digitalization in recent years (Aly 2020; Shahiduzzaman et al. 2018). At the country level, where much of the debate has occurred, economic growth usually refers to the rate of change in GDP (Dedrick et al. 2003). The majority of macro-economic studies have found a favourable association between digitalization and information communication technologies (ICT) and economic growth (Alshubiri et al. 2022), despite specific empirical investigations on this relationship showing a negative result (Krutova et al. 2021). Using the elements of the infrastructure of ICTs and the employment rate of the ICT sector, Aleksandrova and Khabib (2021) illustrated that developed economies are moving towards digitalization more than developed economies during 2019. In contrast, Bahrini and Qaffas (2019) demonstrated a negative impact of ICT on the economic growth of 45 developing countries in Northern African and Western Asia and Sub-Saharan Africa regions over the period 2007-2016, including four ICT proxies such as a number of fixed telephone subscriptions per 100 inhabitants, the number of mobile cellular subscriptions per 100 inhabitants, the number of Internet users per 100 inhabitants and the number of fixed broadband subscriptions per 100 inhabitants. These different results use different arguments about the impact of digital/ICT's influence on productivity and economic growth. Further, Aghaei and Rezagholizadeh (2017) illustrated that there was a significant impact on the rate of investment in IT on both demand and supply functions of an economy from 1990-2014. While there have been numerous studies on the effect of digitalization/ICT on economic growth and productivity in the last years, little has been done in the COVID-19 pandemic and identifying digital/ICT indicators in both market and production digitalization cohorts. The present study can be considered an exception to the rule.

## 2.1 Prominent Contemporary Theories

A significant positive relationship between technology and economic growth has been highlighted by the prominent contemporary theory '*neo-classical growth theory*' (Solow 1956). This theory explains how the interaction of the three forces of labour, capital, and technology leads to a consistent pace of economic growth. Further, this illustrates while an economy's capital and labour resources are limited, the contribution of technology to growth is limitless. Technology thus generates added value at the firm and sectoral levels, improving productivity and fostering economic growth of the country (Dedrick et al. 2003; Fang et al. 2020). While this theory has demonstrated that technology positively affects economic growth, several economists have coined the term '*productivity paradox*' to describe the situation in which productivity growth has slowed due to rapid technological development in the production process (Krutova et al. 2021). For example, Krutova et al. (2021) and Carr (2003) presented this controversial argument on digital adaptation and economic growth and productivity. They further reveal that investments in ICT created a slow labour productivity rate and how to use digitalization prolifically. Likewise, OECD (2019) empirically show the facts of the decline of labour productivity growth in Organization for Economic Cooperation and Development (OECD) countries. Moreover, this puzzling argument has been strengthened by the study based on empirical data presented by Aleksandrova and Khabib (2021), which shows the main differences between developed and developing nations in terms of digital adaptation and Gross Domestic Production (GDP).

## 2.2 Digitalization and Economic Growth: Measures Using ICT/Digital Indicators and Economic Growth Indicators

Digitalization indexes have long been a source of countless openings for personal realization, professional development, and value creation. Technology advances have become vital for working, learning, accessing essential services, and keeping in touch to assess digital activities (Hanafizadeh et al. 2009). In this scenario, plenty of world organizations and scholars kept working and trying to show the potential of ICT in various nations, regions, and the world (Momino and Carrere 2016). International Communication Union (ITU), the World Bank, the Organization for Economic Cooperation and Development (OECD), International Institute for Management Development (IMD), and CISCO are a few examples of organizations that actively participate in this enormous role. Researchers and practitioners have used the digital/ICT/Telecommunication indicators (Aleksandrova and Khabib 2021; Hanafizadeh et al. 2009; Momino and Carrere 2016).

### 2.2.1 Digital/ICT Indicators

**Active Mobile-Broadband Subscriptions (AMBS) (IV1)**: This indicator has been adapted from ITU and used in various other data sources such as CISCO digital readiness index (CISCO 2022), IMD world competitiveness ranking (CISCO 2022), and the world bank database (World Bank 2022). Accordingly, this indicator has been defined as "the sum of active handset-based and computer-based mobile-broadband subscriptions to the public Internet" (ITU 2022a, p. 2). The data has been presented per 100 inhabitants based on data collected annually through questionnaires, government reports, and operators' annual reports. Additionally, Information System (IS) researchers (Aleksandrova and Khabib 2021; Momino and Carrere 2016) have used this indicator to assess international connectivity in communications networks.

**Digital/Technological skills (DS) (IV2):** Under the assessments of the IMD, digital competitiveness ranking has been foreseen to identify digital/technology skills, mobile broadband subscribers, and e-participation of 64 economies (IMD 2021a) with the sub-factor rankings of the regulatory framework, capital, and technology framework (IMD 2021b). IMD (2021b, p.176) defines this as "knowledge and talent necessary to discover, understand and build new technologies". Additionally, this has been taken from IMD World Competitiveness Executive Opinion Survey based on an index from 0 to 10 to see the readily availability. Moreover, the assessment was based on three subcategories; talent, training and education, and scientific concentration (IMD 2021a; IMD 2021b). Besides, Budzinskaya and Teregulova (2021) used digital/technology skills to look at the staffing priorities of a digital economy. Equally, digital skills and knowledge has been identified as one of the influencing factors of the level of digitalization that leads to economic development (Pyroh et al. 2021).

**Fixed Broadband Subscriptions (FBS) (IV3):** World Bank (2022, p.3) has defined the fixed broadband subscriptions indicator as "fixed subscriptions to high-speed access to the public Internet (a TCP/IP connection), at downstream speeds equal to, or greater than, 256 kbit/s". Further, this includes cable modem, DSL, fiber-to-the-home/building, other fixed (wired)-broadband subscriptions, satellite broadband, and terrestrial fixed wireless broadband. Accordingly, this total is measured per 100 people irrespective of the payment method. It excludes subscriptions that have access to data communications (including the Internet) via mobile-cellular networks and includes fixed WiMAX and any other fixed wireless technologies both in residences and organizations (World Bank 2022). Similarly, these world development indicators have been used in previous studies to examine growth and economic development (Aly 2020), Internet usage, and diffusion (Na et al. 2018).

**Households with Internet Access at Home (HIAH) (IV4):** This indicator has been defined as "the proportion of individuals who have used the Internet from home in the last three months that have access to a number of communication services including the world wide web and carries e-mail, news, entertainment, and data files, irrespective of the device used" (ITU 2022b, pp. 1-2). This has been presented as a percentage from 0-100 by ITU to assess international connectivity in communications networks under the key category, ICT infrastructure and access (ITU 2021; World Bank 2022). Moreover, Carroll et al. (2005) and Deursen (2020) have used this in their medical research to assess the digital divide in a paediatric clinic population from the socio-economic status. Besides, Zhang and Maruping (2008) assessed the household technology adoption by incorporating the role of espoused cultural values.

**Individuals Using the Internet (II) (IV5):** As per ITU (2017, p. 48) this illustrates "the proportion of individuals who have used the Internet from any location in the last three months". Internet usage has been assessed as a percentage value from 0 to 100 of the population by looking at access via a fixed or mobile network. Further, this has not been assumed to be only via a computer – it may also be by mobile telephone, tablet, PDA, games machine, digital TV, etc.(ITU 2017, p. 47). This indicator has been used by many researchers and practitioners to look at Internet usage by individuals in a nation. Consequently, different scholars study Internet usage from an individual's perspective to assess the digital inequality (Deursen 2020), diverse outcomes of engaging with the Internet (Deursen and Helsper 2017), and as well as in Internet research using public data sources (Hanafizadeh et al. 2009).

**Mobile Broadband Subscribers (MBS) (IV6):** As per IMD (2021b, p. 176) mobile broadband subscribers have been identified in the "4G & 5G market as a percentage (%) of the mobile market". This has been adopting and exploring digital technologies as a key driver for economic transformation in business, government, and wider society, which has been measured as a percentage of the mobile market on a 0 to 100 scale. 5G connectivity is expected to drive the market growth of the Internet of Things in the coming years, as the newer mobile technology will connect machines and devices with higher data speeds, ultra-low latency, and increased availability, among other benefits. Although 5G offers faster data transfer and lower latency, 4G/LTE will be a major feature of the mobile broadband landscape in the future (Statista 2022). The current scientific literature on the level of digitalization in the world economies combines 4G and 5G mobile markets to understand competitiveness as driving forces to boost economies (Lehr et al. 2021). Equally, it has been widely adapted to recognize advanced communication possibilities promising a growing economy (Schneir et al. 2019).

**Secure Internet Servers (SIS) (IV7):** This indicator has been defined by the World Bank (2022, p. 3) and Netcraft surveys related to "those sites bear the distinct, publicly-trusted TLS/SSL certificate, is valid for the hostname, and the certificate has been issued from a publicly-trusted root in the mean of the use of encrypted transactions through extensive automated exploration" (Indexmundi 2019). Above and beyond the fact that, at the World Bank, Secure Internet Servers have been used to measure the development progress of a nation in relation to science and technology from the Netcraft Secure Server

Survey based on the evaluation of one million people (World Bank 2022). Further, this has been used by scholarly work especially under cyber and information security (Garg et al. 2013) and ethical considerations of web-based behaviour of individuals (Kravchenko et al. 2019).

**Digital Transformation in Companies (DT) (IV8):** In the case of IMD (2021b, p. 32) this has been identified as "the digital transformation takes place at enterprise level (whether private or state-owned) to exploit country's future readiness". This has been ranked on 0-10 scale through effective implementation and satisfying factors of knowledge, technology and future readiness. Knowledge includes the intangible framework that supports the development of new technologies through discovery, application, and learning. Technology is the overall context that allows for the development of digital technologies. The adoption of technology by businesses is examined in terms of readiness. Examples of indicators included in this factor are internet retailing (e-commerce) and data analytics tools in the private sector. Kraft et al. (2022) considered this when conducting their research to evaluate how well SMEs had used digital tools for managerial tasks.

**Use of Big Data and Analytics (BDA) (IV9):** IMD's World Competitiveness Index ranking was considered the use of big data and analytics to look at the digital transformation. As per IMD (2021b, p. 177), the values within 0-10 scale have been presented based on "the companies very good at using big data and analytics to support decision-making" (IMD 2021b). Also, Kirova et al. (2021) have given big data and analytics some thought as a way to institutionalise monitoring the growth of digital economy. Further, the rankings of the IMD Digital Competitiveness Index were also utilised to examine a sustainable and well-balanced digital economy. In order to draw a conclusion from their investigation, the indicators were presented clearly (Morgun et al. 2021).

### 2.2.2 Economic Growth Indicators

**GDP (PPP) per capita (GDPPC) (DV1):** IMD World Competitiveness Index has utilised this to evaluate a nation's domestic economy's performance. This is further defined by IMD (2021b, p.176) as "gross domestic product converted to international dollars using purchasing power parity rates". Using the US dollar per capita at purchasing power parity, this criterion has been estimated (IMD 2021b). Previous studies have been able to take advantage of both better data and larger samples, including time-series data to assess economic performance and economic growth using GDP (PPP) per capita (Alshubiri et al. 2022; Dedrick et al. 2003). Further, Bahrini and Qaffas (2019) were able to take GDP per capita as a proxy for the economic growth of a country.

**GDP Growth (GDPG) (DV2):** As per World Bank (2022, p. 3), GDP has been considered as "the sum of gross value added by all resident producers in the economy plus any product taxes and minus any subsidies not included in the value of the products" in the calculation of GDP growth. This indicator has been calculated by World Bank based on the annual percentage growth rate of GDP at market prices based on constant local currency (World Bank 2022). It is calculated without deductions for depreciation of fabricated assets or depletion and degradation of natural resources (World Bank 2022). GDP growth rate is popular among macroeconomic literature whereas, Peterson (2017) has evaluated the role of population in a country's economic growth by considering the GDP growth factor.

**Gross Domestic Product (GDP) (DV3):** IMD (2021b, p.178) has identified GDP as "a monetary indicator of the total market worth of all the finished products that nations create over a certain time period". As per IMD's Digital Competitiveness Index, GDP has taken to assess the economic performance of a country. Additionally, it contains measurements in a nation's billions of US dollars annually. A study by Demetriades and Hussein (1996) considered GDP to assess the impact of financial development on the economic growth of 16 countries by using time-series evidence. Further, Bahrini and Qaffas (2019), GDP has used this as a key indicator to assess a country's economic growth.

## 3 Methodology

Drawing from previous studies based on diverse streams of ICT/digital indicators and measurements, this study conducts a comprehensive experimental assessment of the association between digitalization and economic growth from the country and regional perspectives using the event study methodology (Nagm and Cecez-Kecmanovic 2009). We have considered COVID-19 pandemic as the event for a period of 3 years from 2018 to 2020, covering the pre-COVID, during, and post-COVID. We have utilised a purposive sampling technique to discover well-known key databases and ICT/digital indicators that are comprised of reliable annual panel data due to the nature and purpose of the study. Panel data comes from the International Telecommunication Union indicators (ITU 2021), the World Development Indicators (World Bank 2022), and IMD World Digital Competitiveness Ranking (IMD 2021b).

## 3.1 Panel Data Estimates

To test our argument that digitalization influences the economic growth of a country, data are used from 59 countries in 7 regions from 2018-2020. Further, we have classified our data into two cohorts as 'HIGH digitalized' and 'LOW digitalized' as per the IMD world digital competitiveness index ranking (IMD 2021b) in the first assessment. The countries lie between 1-32 have been identified as 'HIGH' and 33-64 as 'LOW'. Below are the nations we have considered as the 'HIGH' and 'LOW'.

**HIGH:** Denmark, Norway, Sweden, Finland, Iceland, United Kingdom, Lithuania, Estonia, Ireland, Spain, Belgium, Germany, Austria, Netherlands, Luxembourg, Switzerland, France, Canada, Korea Rep., Japan, China, Singapore, Kazakhstan, Australia, New Zealand, Israel, Qatar
**LOW:** Latvia, Croatia, Greece, Italy, Portugal, Bulgaria, Slovenia, Hungary, Poland, Romania, Russia, Czech Republic, Slovak Republic, Ukraine, Venezuela, Colombia, Mexico, Argentina, Chile, Brazil, Peru, Mongolia, Thailand, Indonesia, Philippines, India, Turkey, Saudi Arabia, Cyprus, Jordan, Botswana, South Africa

The dependent variable (DV) is economic growth, and the key independent variable (IV) is digitalization. Based on the comprehensive data availability, we have identified 9 significant digital/ICT indicators[1] to investigate the IV and 3 economic growth indicators[2] to assess the DV. Further, blending the adaptation of digitalization with digitalization indexes, we have further categorized 9 digital/ICT indicators as market[3] and production[4] indicators. All indicators were analyzed by 7 regions across different countries, similar to the studies that have been used with similar criteria (Aleksandrova and Khabib 2021; Bahrini and Qaffas 2019; Shahiduzzaman et al. 2018). The regional assessment was considered the second assessment of the data analysis. The regions are Europe and Northern America, Latin American And Caribbean, Eastern and South-Eastern Asia, Central and Southern Asia, Oceania, Northern Africa and Western Asia, and Sub- Saharan Africa. As per the United Nations' categorizations (Beltekian 2021; United Nations 2021; United Nations 2022), the regions and particular countries we have considered for this study assessment are shown below. Additionally, countries were chosen based on the requirement that at least 2 data points for each of the seven indicators be present.

**Europe and Northern America:** Denmark, Norway, Sweden, Finland, Iceland, United Kingdom, Lithuania, Latvia, Estonia, Ireland, Croatia, Greece, Italy, Spain, Portugal, Slovenia, Hungary, Poland, Romania, Bulgaria, Russia, Czech Republic, Slovak Republic, Ukraine, Belgium, Germany, Austria, Netherlands, Luxembourg, Switzerland, France, Canada
**Latin American and Caribbean:** Venezuela, Argentina, Brazil, Colombia, Chile, Peru, Mexico
**Eastern and South-Eastern Asia:** Korea Rep., Mongolia, Japan, China, Singapore, Indonesia, Philippines, Thailand
**Central and Southern Asia:** Kazakhstan, India
**Oceania:** Australia, New Zealand
**Northern Africa and Western Asia:** Turkey, Cyprus, Israel, Qatar, Jordan, Saudi Arabia
**Sub- Saharan Africa:** Botswana, South Africa

To do this, we have developed the equations below to calculate digitalization and economic growth. Equation (1) is equivalent to the calculation of digitalization, and equation (2) is about the calculation of the economic growth of a country. We used Bahrini and Qaffas (2019) and Ward and Zheng (2016) studies as grounded work and presented the equations accordingly, where i represents each country in the panel and t indicates the time period[5].

$Dig._{it} = \sum + \beta_1 (AMBS) + \beta_2 (DS) + \beta_3 (FBS) + \beta_4 (HIAH) + \beta_5 (II) + \beta_6 (MBS) + \beta_7 (SIS) + \beta_8 (DT) + \beta_9 (BDA)$ **(1)**

$EG._{it} = \sum + \beta_1 (GDPPC) + \beta_2 (GDPG) + \beta_3 (GDP)$ **(2)**

# 4 Results and Discussion

The results are provided from two contexts: digitalized country classification as "HIGH" or "LOW," and regional study.

## 4.1 The Analysis of the Relationship Between Digitalization and Economic Growth in the 'HIGH' and 'LOW' Digitalized Countries

Data from Figure 1 (a), scatter plot diagram illustrates positive correlation between digitalization and economic growth in the 'HIGH' digitalized countries (Pearson Correlation= 0.393; Sig. (2-tailed) =0.000), so digitalization and economic growth increased in the 2018-2020. However, as per sequential diagram, Figure 1 (b), digitalization increased slower than economic growth.

---

[1] IV1: Active mobile broadband subscriptions (per 100 inhabitants), IV2: Digital/ Technology Skills (readily available), IV3: Fixed broadband subscriptions (per 100 people), IV4: Households with Internet access at home (% of population), IV5: Individuals using the Internet (% of population), IV6: Mobile broadband subscriptions- 4G & 5G market (% of mobile market), IV7: Secure Internet servers (per 1 million people), IV8: Digital transformation in companies (WCY), IV9: Use of big data and analytics (WCY)
[2] DV1: GDP (PPP) per capita (WCY)-US$ per capita at purchasing power parity, DV2: GDP growth (annual %), DV3: Gross Domestic Product (GDP) (WCY)-US$ billions
[3] IV1-IV6
[4] IV7-IV9
[5] The detailed calculation table can be received upon request

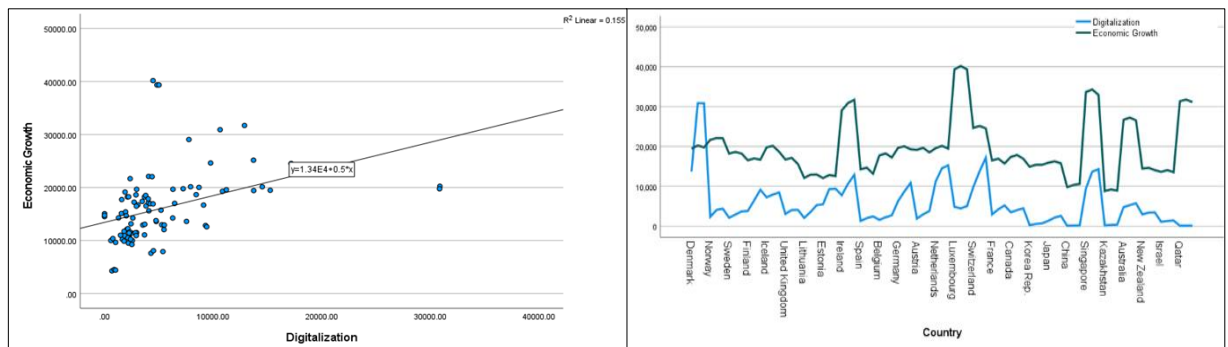

*Figure 1 (a,b): Correlation of Digitalization and Economic Growth in the 'High' Digitalized Countries*

The scatter plot diagram in Figure 2 (a) depicts positive correlation between digitalization and economic growth in the 'LOW' digitalized countries (Pearson Correlation = 0.468; Sig. (2-tailed) =0.000). It is also evident that the correlation is higher in "LOW" digitalized countries than "HIGH" digitalized countries. Though both digitalization and economic growth expanded, however, sequential diagram, Figure 2 (b) shows digitalization increased more slowly than economic growth.

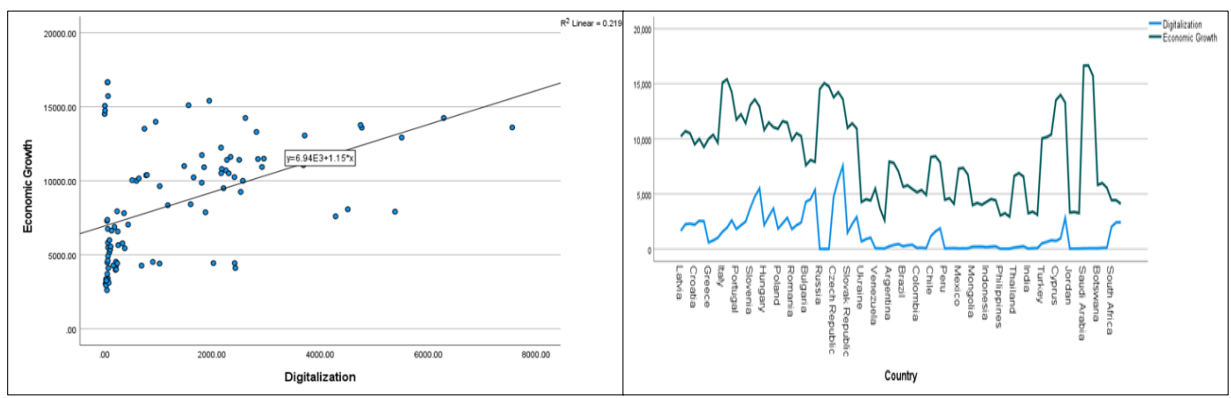

*Figure 2 (a,b): Correlation of Digitalization and Economic Growth in the 'LOW' Digitalized Countries*

Further, we analysed the impact of digitalization on economic growth in both 'HIGH' and 'LOW' digitalized countries and presented the results in Table 1. As per the Table, the estimated simple regression model for the impact of digitalization on economic growth in 'High' digitalized economies, highlights that digitalization positively influenced economic growth (β=0.292). This model was statistically significant (t=2.709; p=0.008) and accounted for 7.3% of the variance of economic growth (Adjusted $R^2$=0.073). The results of 'LOW' digitalized countries also highlight that digitalization positively influences economic growth (β=0.468) and the model was statistically significant (t=5.147; p=0.000) and accounted for 21.1% of the variance of economic growth (Adjusted $R^2$=0.211).

|  | Correlation |  | Coefficients |  |  | Model Summary |  |
|---|---|---|---|---|---|---|---|
|  | Pearson Correlation | Sig. | β | t | Sig. | Adjusted $R^2$ | Sig. |
| 'HIGH' Countries | 0.393 | 0.000 | 0.292 | 2.709 | 0.008 | 0.073 | 0.008 |
| 'LOW' Countries | 0.468 | 0.000 | 0.468 | 5.141 | 0.000 | 0.211 | 0.000 |

*Table 1. Regression of Digitalization on Economic Growth of 'HIGH' and 'LOW' Digitalized countries*

Though the impact of digitalization on economic growth in 'LOW' digitalized economies is higher, the variation of economic growth lies between 18,000-2,500 and digitalization in 8,000-0 range (Figure 2 (b)). In 'HIGH' digitalized countries, economic growth has lied between 40,000-9,500 range and digitalization in 31,000-0 range (Figure 1 (b)). According to the United Nations' Human Development Index (HDI), the majority of "HIGH" digitalized countries are industrialised countries while "LOW" digitalized countries are developing countries (UNDP 2020). In order to extend and integrate digital technology into their socio-economic processes, developed countries are moving towards digitalization. The primary purposes of using digital technologies in developing countries are to promote the development and operation of consumer markets and to meet the informational and communicational requirements of consumers (Aleksandrova and Khabib 2021). Furthermore, in "HIGH" digitalized countries, the impact that digitalization can make is relatively modest, whereas in "LOW" digitalized countries, the digitalization can make a greater impact (Bahrini and Qaffas 2019).

### 4.2 The Analysis of the Relationship Between Digitalization and Economic Growth as per Region

We have used correlation and regression analysis as well as sequential diagrams to determine the effects and behavioural patterns of digitalization and economic growth in different regions between 2018–2020. According to the panel data acquired, the correlation matrix and regression findings are shown in Table 2. As per the correlation values, it is evident that Eastern and South-Eastern Asia (Pearson Correlation= 0.898; Sig. (2-tailed) =0.000), Central and Southern Asia (Pearson Correlation= 0.874; Sig. (2-tailed) =0.011) and Oceania (Pearson Correlation= 0.947; Sig. (2-tailed) =0.002) regions had a very strong positive correlations between digitalization and economic growth. Similarly, the Latin American and Caribbean region had a positive, much higher positive correlation (Pearson Correlation= 0.609; Sig. (2-tailed) = 0.003) between digitalization and economic growth. There was a positive correlation in Europe and North America, the relationship was particularly significant (Pearson Correlation= 0.393; Sig. (2-tailed) =0.000). Accordingly, digitalization has become an integral part of economic activities in most of the developed countries lying within these regions (Aleksandrova and Khabib 2021). Conversely, Northern Africa and Western Asia (Pearson Correlation= -0.137; Sig. (2-tailed) =0.294) and Sub-Saharan Africa (Pearson Correlation= -0.979; Sig. (2-tailed) = 0.000) had negative correlation between digitalization and economic growth. As per the Table 2, our findings show insignificant correlation between digitalization and economic growth for Northern Africa and Western Asia (sig < 0.05). Furthermore, we analysed simple regression for the impact of digitalization on economic growth region wise and presented results in Table 2 As per the Table, it is evident that digitalization positively influenced economic growth in all regions except Northern African and Western Asia and Sub-Saharan Africa ($\beta$=-0.137, $\beta$=-0.979, respectively). Europe and Northern America's model was statistically significant (t=4.147; p=0.000) and accounted for 14.6% of the variance of economic growth (Adjusted $R^2$=0.146). Latin American and Caribbean represent 33.8% of the variance (Adjusted $R^2$= 0.338) of the economic growth with statistically significant model (t=3.346; p=0.000). Similarly, Eastern and South-Eastern Asia with 79.8% (Adjusted $R^2$= 0.798; t=9.576; p=0.000), Central and Southern Asia with 70.4% (Adjusted $R^2$= 0.704; t=3.591; p=0.023), and Oceania with 87.2% (Adjusted $R^2$= 0.872; t=5.920; p=0.004) (Table 2 highlights certain items in red). As per the table, there was insignificant relation between digitalization on economic growth in Northern African and Western Asia.

|  | Correlation | | Coefficients | | | Model Summary | |
|---|---|---|---|---|---|---|---|
|  | Pearson Correlation | Sig. | β | t | Sig. | Adjusted $R^2$ | Sig. |
| Europe and Northern America | 0.393 | 0.000 | 0.393 | 4.147 | 0.000 | 0.146 | 0.000 |
| Latin American and Caribbean | 0.609 | 0.003 | 0.609 | 3.346 | 0.000 | 0.338 | 0.003 |
| Eastern and South-Eastern Asia | 0.898 | 0.000 | 0.898 | 9.576 | 0.000 | 0.798 | 0.000 |
| Central and Southern Asia | 0.874 | 0.011 | 0.874 | 3.591 | 0.023 | 0.704 | 0.023 |
| Oceania | 0.947 | 0.002 | 0.947 | 5.920 | 0.004 | 0.872 | 0.004 |
| Northern Africa and Western Asia | -0.137 | 0.294 | -0.137 | -0.553 | 0.588 | -0.043 | 0.588 |
| Sub-Saharan Africa | -0.979 | 0.000 | -0.979 | -9.507 | 0.001 | 0.947 | 0.001 |

*Table 2. Regional Analysis: Simple Regression Results*

As per Tables 2, it is clear that Northern Africa and Western Asia, and Sub-Saharan Africa have slightly different effects of digitalization on economic growth and behavioural patterns than the other regions. Therefore, we combined 'HIGH' and 'LOW' digitalized countries with regional evaluation for further examination of this peculiar pattern using sequential diagrams. Figure 3 shows the behavioural trends associated with digitalization and economic expansion in the two regions. We performed a basic regression analysis to examine each digital/ICT indicator's (IVs) impact on economic growth in order to delve deeper into the aforementioned behaviour of Northern Africa and Western Asia and Sub-Saharan Africa Regions. Our results further illustrate that, use of big data and analytics (I9) was the only driver of economic growth in Northern Africa and Western Asia region ($\beta$=3.217, Sig.=0.024). Even though IV5, Individuals using the Internet, shows a positive association, the result was not statistically significant ($\beta$=1.167, Sig.=0.296). All the other indicators negatively impacted the economic growth over the recent period 2018-2020 in Northern Africa and Western Asia. In fact, the negative consequences of digitalization on economic growth highlighted that many developing countries in our sample did not leverage the growth potential of this adoption of digital technology. This could be because they are unable to afford investment needed to extend digital infrastructure across their borders (Wamboye et al. 2015). Unfortunately, the absence of data in the digital/ICT variables prevented us from a simple regression analysis for the Sub-Saharan Africa region.

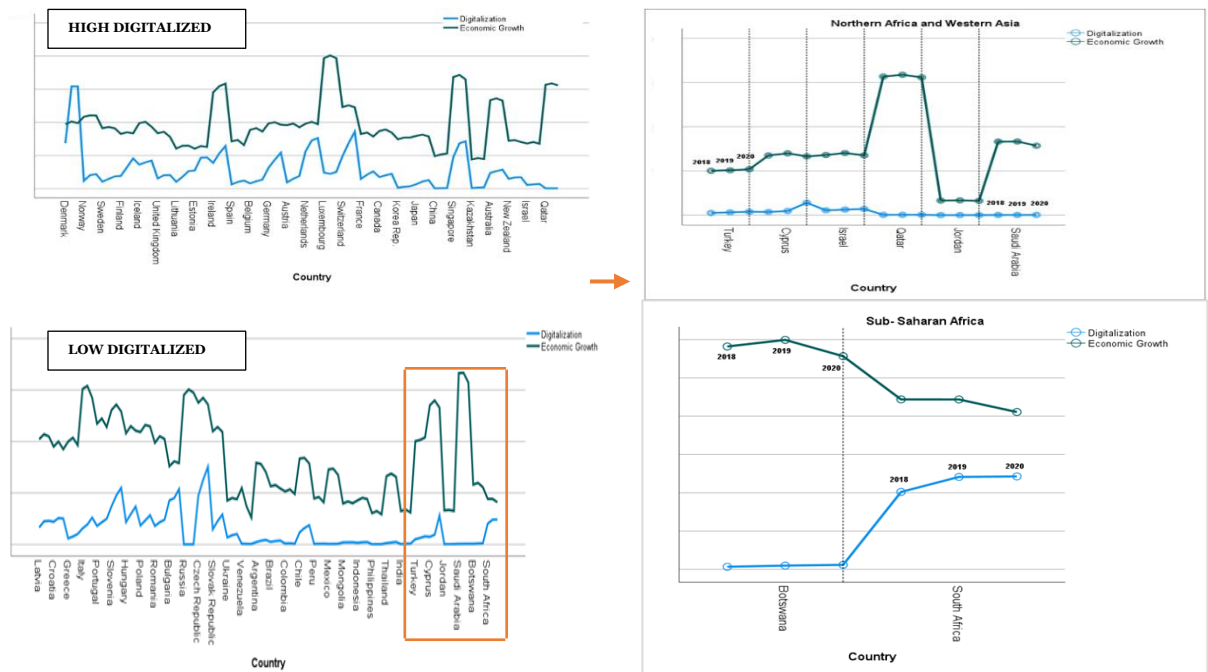

*Figure 3: Behavioural Patterns of IV and DV, Combined Effect of 'HIGH', 'LOW' and Regions*

## 5　Conclusion, Contributions and Future Research

In this study, we investigated the effect of digitalization on the economic growth of 59 countries from the 7 regions over the period 2018–2020. A correlation matrix and simple regression along with sequential diagrams were employed to explore the nexus between 3 economic growth indicators and 9 digital/ICT indicators. Findings show a positive correlation between digitalization and economic growth in both 'HIGH' and 'LOW' digitalized countries. In accordance with earlier studies, this confirms that digitalization and technological developments are the key drivers of economic growth (Kraft et al. 2022; Kravchenko et al. 2019). However, digitalization has increased slower than the economic growth in both 'HIGH' and 'LOW' digitalized countries. In contrast to 'HIGH' digitalized countries, the correlation results are more robust in 'LOW' digitalized countries. Regression analysis reveals a comparable finding, demonstrating a strong statistically significant positive variance of economic growth in countries with "LOW" levels of digitalization. Indeed, the results of correlation and regression analysis of 'HIGH' digitalized countries reveal that many countries in our sample had given little attention to benefit from the growth potential of these digital/ICT indicators. As per the regional analysis, all the regions except Northern Africa and Western Asia, and Sub-Saharan Africa Regions, show a positive correlation and statistically significant regression results between digitalization and economic growth. This finding is consistent with previous studies used in cross-country analysis (Aleksandrova and Khabib 2021; Bahrini and Qaffas 2019). Further, it is evident from the close examination of the behavioural trends of digitalization and economic growth in Northern Africa and Western Asia, and Sub-Saharan Africa that the use of big data and analytics (I9) was the only driver of economic growth. The findings can contribute to both IT investment and economic applications. The primary theoretical contribution of this study is that it provides insights into widely used theories and concepts of neo-classical growth theory and productivity paradox for digital adaptation and economic performance. This study has also significant practical implications. Our findings recommend that authorities and policymakers expand investment in digital technologies to promote sustainable economic growth. We acknowledge that 9 digital/ICT and 3 economic growth indicators may be limited in their scope, and may not capture the full magnitude of digitalization and economic growth of a country. There may be nuanced indicators of digitalization, economic growth and geographical sensitivity that might further highlight digitalization trends and economic growth. Therefore, a future study could include further factors to understand more precisely what makes/not make nations adopt more digital technologies, during the COVID-19 pandemic along with the economic growth. Further, an analysis can be presented through a separate indicator assessment based on the market and production digitalization indicators.